\documentclass[aps,prl,twocolumn]{revtex4}
\usepackage{graphicx,epsf}
\usepackage{amsmath,amssymb}

\usepackage{color}

\newcommand{\tlcu}{TlCuCl$_3$}
\newcommand{\bacu}{BaCuSi$_2$O$_6$}
\newcommand{\lsco}{La$_{2-x}$Sr$_x$CuO$_4$}

\newcommand{\w}{\omega}

\newcommand{\Nr}{N_{\rm r}}

\begin{document}

\title{
%Excitations of disordered dimer magnets near quantum criticality:\\The spectrum of a Griffiths phase
Excitation spectra of disordered dimer magnets near quantum criticality
}

\author{Matthias Vojta}
\affiliation{Institut f\"ur Theoretische Physik, Technische Universit\"at Dresden,
01062 Dresden, Germany}
\date{Aug 16, 2013}

\begin{abstract}
For coupled-dimer magnets with quenched disorder, we introduce a generalization of the
bond-operator method, appropriate to describe both singlet and magnetically ordered
phases. This allows for a numerical calculation of the magnetic excitations at all
energies across the phase diagram, including the strongly inhomogeneous Griffiths regime
near quantum criticality. We apply the method to the bilayer Heisenberg model with bond
randomness and characterize both the broadening of excitations and the transfer of
spectral weight induced by disorder. Inside the antiferromagnetic phase this model
features the remarkable combination of sharp magnetic Bragg peaks and broad magnons, the
latter arising from the tendency to localization of low-energy excitations.
\end{abstract}
\pacs{74.72.-h,74.20.Mn}

\maketitle

%%%%%%%%%%%%%%%%%%%%%%%%%%%%%%%%%%%%%%%%%%%%%%%%%%%%%%%%%%%%%%%%%%%%%%%

Magnetic quantum phase transitions (QPT) have attracted enormous interest over the past two
decades, with intriguing aspects such as Bose-Einstein condensation of magnons, exotic
criticality, and non-Fermi liquid behavior
%, and secondary instabilities, e.g., towards superconductivity
\cite{ssbook,ss_natph,gia08,hvl}.
The influence of quenched disorder, being inevitable in condensed-matter systems, on QPTs
is a less explored field, although a number of theoretical results are
available \cite{tv06,tv10}: Disorder can modify the critical behavior or even destroy the QPT; in addition
it can produce singular response in off-critical systems via quantum
Griffiths physics dominated by rare regions \cite{griffiths}.

While the thermodynamics of disordered model systems near quantum criticality has been
studied using a variety of theoretical tools \cite{tv06}, rather little is known about the
dynamics of excitations in this fascinating regime, mainly because numerical methods are
either restricted to the ground state or work in imaginary time where real-frequency
spectra are difficult to extract, while analytical methods are restricted to low energy
and to the limits of either weak or strong disorder.
For quantum magnets, this issue is pressing, as high-resolution inelastic neutron
scattering (INS) experiments are getting access to magnetic excitations near QPTs
\cite{ruegg04,lake05,ruegg08,chang09}, and suitable materials with disorder created via intentional
doping are available \cite{stone10,zheludev12,zheludev12b,keimer12}.

The aim of this paper is to close this gap:
For coupled-dimer magnets, being paradigmatic model systems for magnetic QPTs
\cite{ssbook,ss_natph,gia08}, we propose a generalization of the bond-operator method
\cite{bondop} to cases with quenched disorder. This enables the numerical calculation of
magnetic excitation spectra in both paramagnetic and magnetically ordered phases,
including the vicinity of the QPT, with disorder being treated exactly in finite-size
systems.
We apply the method to the square-lattice bilayer Heisenberg model, studied in detail in
the disorder-free case \cite{hida90,hida92,SaSca,Chub95,kotov,sandvik06}, with different types of
exchange randomness. Near quantum criticality, we generically find a strong broadening in
both energy and momentum of the low-energy response, while that at high energy
is less affected by disorder.
An interesting dichotomy follows upon slightly moving into the antiferromagnetic (AF)
phase: the strongly inhomogeneous AF features low-energy magnons that are distinctly
broadened due to disorder-driven mode localization, but at the same time displays sharp
Bragg peaks in its elastic response.
% (Well-defined Goldstone modes emerge only at extremely small energies.)
We discuss connections to recent INS experiments.
% as well as other applications of our novel approach.

%%%%%%%%%%%%%%%%%%%%%%%%%%%%%%%%%%%%%%%%%%%%%%%%%%%%%%%%%%%%%%%%%%%%%%%

{\it Model.}
We consider a Heisenberg magnet of coupled pairs of spins 1/2,
with the general Hamiltonian
\begin{equation}
\label{h}
\mathcal{H} =
\sum_i J_i \vec{S}_{i1} \cdot \vec{S}_{i2} +
\sum_{\langle ii'\rangle mm'} K_{ii'}^{mm'} \vec{S}_{im} \cdot \vec{S}_{i'm'}
\end{equation}
defined on a lattice of $N$ dimer sites $i$. $J$ and $K$ are the intradimer and interdimer
couplings, and $m=1,2$ labels the two spins of each dimer.
Without quenched disorder in the couplings $J$ and $K$, this Hamiltonian can describe,
e.g., spin ladders and bilayer Heisenberg magnets, but also applies to materials like
\tlcu\ \cite{ruegg04,ruegg08,tlcucl,oosawa99,rueggtlcucl,MatsumotoNormandPRL} and
\bacu\ \cite{sasago,bacusio,sebastian}.

%%%%%%%%%%%%%%%%%%%%%%%%%%%%%%%%%%%%%%%%%%%%%%%%%%%%%%%%%%%%%%%%%%%%%%%

{\it Generalized bond-operator method.}
Let the four states of each dimer $i$ be $|t_k\rangle_i$,
$k=0, \ldots, 3$, where
$|t_0\rangle=
(|\uparrow\downarrow\rangle -|\downarrow\uparrow\rangle)/\sqrt{2}$,
$|t_1\rangle =
(-|\uparrow\uparrow\rangle+|\downarrow\downarrow\rangle)/\sqrt{2}$,
$|t_2\rangle =
i(|\uparrow\uparrow\rangle+\downarrow\downarrow\rangle)/\sqrt{2}$,
$|t_3\rangle =
(|\uparrow\downarrow\rangle+|\downarrow\uparrow\rangle)/\sqrt{2}$.
Formally, bosonic operators $t_{ik}^\dagger$ can be introduced which create
these states out of a fictituous vacuum, $|t_k\rangle_i = t_{ik}^\dagger |vac\rangle_i$,
with the constraint $\sum_k  t^\dagger_{ik} t_{ik} =1$ defining the physical Hilbert
space \cite{bondop}.
In a paramagnetic phase, it is convenient to fully ``condense'' the singlet,
such that the operators $t_{i\alpha}^\dagger$ ($\alpha=1,2,3$)
now create triplet excitations (``triplons'') from a singlet background,
and the constraint becomes of hard-core type,
$\sum_\alpha  t^\dagger_{i\alpha} t_{i\alpha} \leq 1$ \cite{kotov}.
Approximations can then be understood as expansion about the singlet
product state $|\psi_0\rangle = \prod_i |t_0\rangle_i$.

%%%%%%%%%%%%%%%%%%%%%%%%%%%%%%%%%%%%%%%%%%%%%%%%%%%%%%%%%%%%%%%%%%%%%%%

\begin{figure*}[!t] % FIGURE 1
\includegraphics[width=0.91\textwidth]{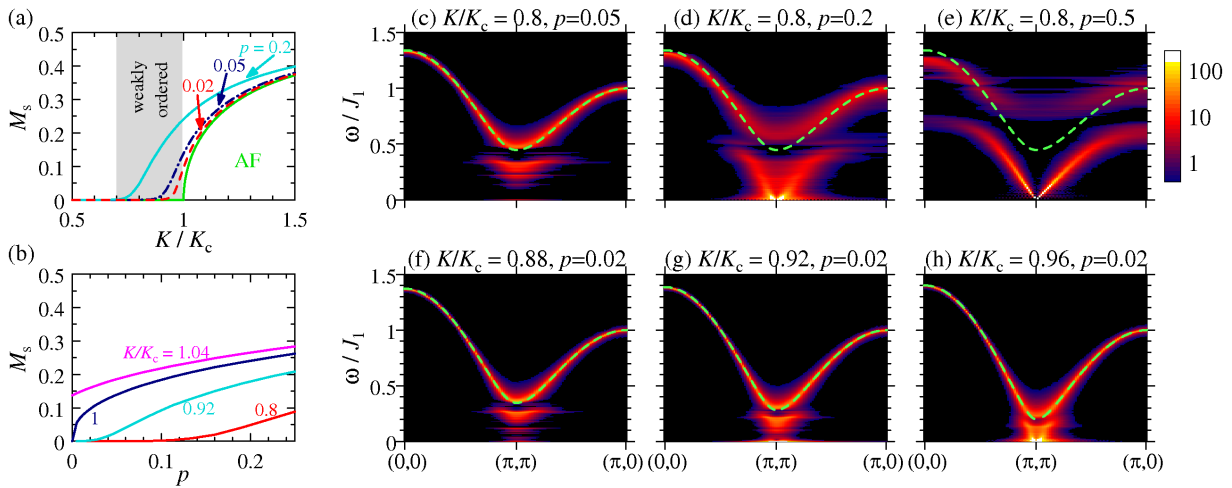}
\caption{(color online)
Numerical results for the disordered bilayer Heisenberg model, with a
concentration of $p$ weak interlayer couplings with $J_2 = J_1/2$.
a) Phase diagram, showing the staggered magnetization $M_s$ as function of $K/K_c$ for
different levels of disorder $p$, with $K_c = J_1/4$ denoting the critical coupling of
the clean system in the linearized bond-operator approach. A regime with weak and
strongly inhomogeneous order emerges for small $p$ and $K\lesssim K_c$.
b) $M_s$ as function of doping level $p$ for different $K$.
c-h) Transverse dynamic susceptibility $\chi''(\vec{q},\w)$ along a path in the 2d
Brillouin zone ($q_z\!=\!\pi$) for different combinations of $K/K_c$ and $p$. The
green dashed lines indicate the dispersion of the clean system with the respective $K$.
} \label{fig:spec1}
\end{figure*}

%%%%%%%%%%%%%%%%%%%%%%%%%%%%%%%%%%%%%%%%%%%%%%%%%%%%%%%%%%%%%%%%%%%%%%%

Magnetically ordered phases correspond to a condensate of triplets, such that the
reference state now involves a linear combination of singlet and triplets on each site. A
consistent description of excitations requires a basis rotation in the four-dimensional
Hilbert space spanned by the $|t_k\rangle$ \cite{sommer,penc}. Here we generalize the approach
of Ref.~\onlinecite{sommer} to inhomogeneous states.
For every dimer site, we introduce a SU(4) rotation to new basis
states,
\begin{equation}
\label{trafo}
|\tilde{t}_k\rangle_i = U_{kk'}^{(i)} |t_{k'}\rangle_i,~
\tilde{t}_{ik}^\dagger = U_{kk'}^{(i)} t_{ik'}^\dagger
~~~~(k,k'=0,\ldots,3)
\end{equation}
such that $|\tilde{\psi}_0\rangle = \prod_i |\tilde{t}_0\rangle_i$ replaces
$|\psi_0\rangle$ as the reference state. For instance,  $|\tilde{t}_0\rangle =
(|t_0\rangle + |t_3\rangle)/\sqrt{2} = |\uparrow\downarrow\rangle$ describes a N\'eel state
polarized along $z$.
The $U^{(i)}$ are chosen such that $|\tilde{\psi}_0\rangle$ is the best product-state
(i.e. saddle-point) approximation to the ground state of $\mathcal{H}$ \cite{suppl}.

One now re-writes the Hamiltonian \eqref{h} in terms of the $|\tilde{t}_{k}\rangle_i$
using the transformation \eqref{trafo} \cite{suppl}.
In analogy to the paramagnetic case, one condenses $\tilde{t}_0^\dagger$, such that
the $\tilde{t}_\alpha^\dagger$ ($\alpha=1,2,3$) describe excitations on top of the
reference state $|\tilde{\psi}_0\rangle$. The Hamiltonian takes the form
$\mathcal{H} = \mathcal{H}_0 + \mathcal{H}_1 + \mathcal{H}_2 + \mathcal{H}_3 +
\mathcal{H}_4$, where $\mathcal{H}_n$ contains $n$ $\tilde{t}_\alpha^{(\dagger)}$ % ($\alpha=1,2,3$)
operators, and an additional hard-core constraint for the $\tilde{t}_\alpha$ is implied.
With the proper (saddle-point) choice of the reference state, $\mathcal{H}_1$ vanishes,
and $\mathcal{H}_2$ describes Gaussian fluctuations around an inhomogeneous magnetic
state. $\mathcal{H}_2$ has the form
\begin{equation}
\mathcal{H}_2 = \sum_{ij\alpha}
\left[
A_{ij} \tilde{t}_{i\alpha}^\dagger \tilde{t}_{j\alpha} + \big(\frac{B_{ij}}{2} \tilde{t}_{i\alpha} \tilde{t}_{j\alpha} + h.c.\big)
\right]
\end{equation}
where quenched disorder enters via random $A_{ij}$, $B_{ij}$.
$\mathcal{H}_2$ is solved by a bosonic Bogoliubov transformation, which yields the $3N$
positive-energy eigenmodes of $\mathcal{H}_2$, used to calculate the dynamic spin
susceptbility \cite{suppl}. In both paramagnetic and collinearly ordered phases, the
polarization directions decouple, giving rise to triply degenerate modes in the
paramagnetic case and a longitudinal and two transverse modes in the collinear case.

The excitations described by this method interpolate continuously between triplons of a
paramagnet and spin waves of a semiclassical AF \cite{sommer,mvtu04}.
The present harmonic approximation constitutes the leading-order correction to
$|\tilde\psi_0\rangle$ in a systematic expansion in $1/z$ with $z$ the
number of neighbors \cite{suppl,joshi}. Anharmonic effects are neglected at this level;
near criticality, this is qualitatively permissible if the anomalous exponent $\eta$ is
small [$\eta=0.03$ for the Heisenberg model in $(2+1)$ dimensions]. Their quantitative
effect on the dispersion can be captured via a renormalization of model parameters
\cite{eder98}; below we account for this by specifying parameters relative to the
location of the QPT.

%%%%%%%%%%%%%%%%%%%%%%%%%%%%%%%%%%%%%%%%%%%%%%%%%%%%%%%%%%%%%%%%%%%%%%%

{\it Bilayer magnet.}
In the remainder of the paper, we illustrate the application the method to a simple case,
namely the bilayer Heisenberg model, as realized e.g. in \bacu\ \cite{sasago,bacusio,sebastian}.
Here, the dimers live on a 2d square lattice with interlayer coupling $J$ and intralayer
coupling $K=K_{ii'}^{11}=K_{ii'}^{22}$ for nearest-neighbor sites $i,i'$. In the absence
of disorder, this model is in a singlet ground state for $J\gg K$ and an AF
ground state with ordering wavevector $\vec{Q}=(\pi,\pi)$ for $J\ll K$, with a O(3)
critical point at $(J/K)_c = 2.5220(1)$ \cite{sandvik06}.
In the harmonic bond-operator approach, this transition occurs at $(J/K)_c=4$; including
triplon interactions allows one to obtain a value very close to the exact
result \cite{kotov}.

With disorder of random-mass type, the character of the QPT changes, as
dictated by the Harris criterion \cite{harris}.
% $\nu>2/d$ required for stability.
Numerical simulations have shown that a new critical point with conventional power-law
singularities obtains \cite{sknepnek} for not too strong disorder \cite{igloi06}.
Here, we shall focus on the excitation spectrum in the vicinity of this QPT
upon inclusion of disorder where signatures of quantum Griffiths behavior can be
expected \cite{tv06,tv10}.

%%%%%%%%%%%%%%%%%%%%%%%%%%%%%%%%%%%%%%%%%%%%%%%%%%%%%%%%%%%%%%%%%%%%%%%

{\it Modelling disorder.}
We shall mainly employ bimodal distributions of coupling constants, being experimentally
relevant to cases where chemical substitution modifies exchange paths, as occurs, e.g.,
in Tl$_{1-x}$K$_x$CuCl$_3$ \cite{tlcudis} or (C$_4$H$_{12}$N$_2$)Cu$_2$(Cl$_{1-x}$Br$_x$)$_6$
\cite{zheludev12b}.
Provided that all couplings remain AF, this type of bond randomness
does not introduce frustration, such that the magnetic order realized for $J\ll K$
continues to be collinear with wavevector $\vec{Q}=(\pi,\pi)$. We will denote the
corresponding order parameter, the staggered magnetization per spin, by $M_s$.

%%%%%%%%%%%%%%%%%%%%%%%%%%%%%%%%%%%%%%%%%%%%%%%%%%%%%%%%%%%%%%%%%%%%%%%

{\it Results: Weak intradimer bonds and evolution across QPT.}
We now present numerical results of our approach, for disordered intradimer coupling
$J$, with values $J_1$ and $J_2$ taken with probabilities $(1-p)$ and $p$, respectively,
and use the interdimer coupling $K$ as a tuning parameter to access the transition.
Energies will be quoted in units of $J_1$, the system size is $N=L^2$ with $L=64$ and
additional $2^2$ supercells \cite{suppl}, disorder averages are performed over $\Nr=50$
realizations, and the temperature is $T=0$.% unless noted otherwise.

Results for small concentrations $p$ of weak intradimer bonds, $J_2=J_1/2$, are
in Fig.~\ref{fig:spec1}:
Panels (a,b) display $M_s$ as function of $K$ and $p$, while panels (c)--(h) show the
dynamical susceptibility $\chi''(\vec{q},\w)$ for different parameter sets.
This type of substitution drives the system towards the ordered phase, by decreasing the
apparent gap and shifting the QPT to smaller $K$ upon increasing $p$. There is a broad
range of parameters with weak magnetic order, i.e., small non-zero $M_s$, to be discussed
below.

\begin{figure}[!b]
\includegraphics[width=0.49\textwidth]{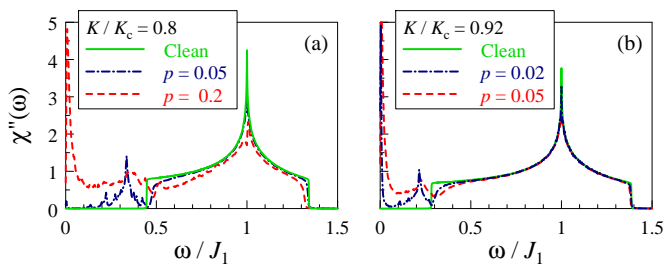}
\caption{(color online)
Momentum-integrated transverse susceptibility $\chi''(\w)$, for the bilayer
model with random weak interlayer couplings as in Fig.~\ref{fig:spec1}.
Disorder tends to close the spin gap by transferring of spectral weight to low energies.
}
\label{fig:sw}
\end{figure}

\begin{figure}[!t]
\includegraphics[width=0.49\textwidth]{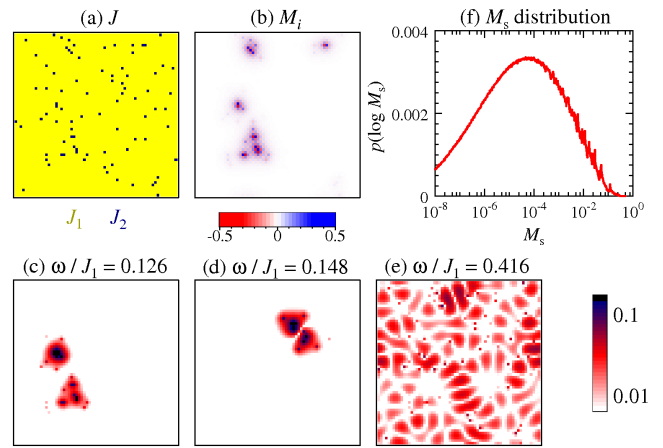}
\caption{(color online)
Bilayer Heisenberg magnet with $K/K_c=0.92$ and a single disorder realization of $p=0.02$
weak intradimer bonds with $J_2=J_1/2$, as in Fig.~\ref{fig:spec1}(g), with $L=64$.
(a) Spatial distribution of $J$ values, indicating the location of the weak bonds.
(b) Corresponding spatial distribution of the local magnetization $M_i$.
(c)-(e) Wavefunction amplitudes of selected eigenmodes of $\mathcal{H}_2$.
(f) Disorder-averaged probability distribution of local (staggered) magnetization values
for parameters as in (a)-(e).
}
\label{fig:magwf}
\end{figure}

Introducing a small amount $p$ of disorder into the paramagnetic clean
system causes a transfer of spectral weight to low energies, Fig.~\ref{fig:spec1}(c,f-h) and
Fig.~\ref{fig:sw}: Doping creates excitations inside the gap which eventually induce weak
order upon increasing $p$.
This disorder-induced low-energy response, although centered around the
ordering wavevector $\vec{Q}$, is broad in both momentum and energy (except for
Goldstone modes at extremely small $\w$): This reflects the localization tendency of
the corresponding modes, see Fig.~\ref{fig:magwf}.
For intermediate $p$ near 1/2, the spectrum visibly separates into upper and lower
branches which exist over the entire Brillouin zone, Fig.~\ref{fig:spec1}(e); these
branches correspond to modes which are primarily carried by either the $J_1$ or the $J_2$
intradimer bonds.

\begin{figure}[!b]
\includegraphics[width=0.49\textwidth]{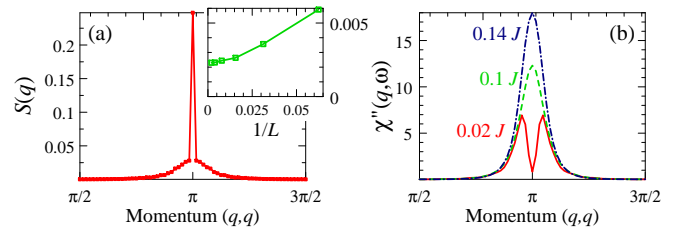}
\caption{(color online)
(a) Static structure factor $S(\vec{q})$ and
(b) susceptibility $\chi''(\vec{q},\w)$ at different $\w$,
for $K/K_c=0.92$ and $p=0.02$ weak intradimer bonds for $L=128$.
The inset in (a) shows the finite-size scaling of $[S(\vec{Q})/(2N)]^{1/2}$
which equals the order parameter $M_s$ in the thermodynamic limit.
}
\label{fig:stfac1}
\end{figure}

\begin{figure*}[!t]
\includegraphics[width=0.9\textwidth]{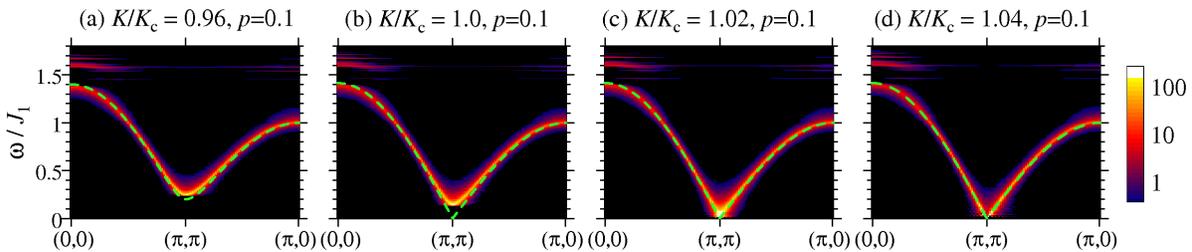}
\caption{%(color online)
Results for the susceptibility $\chi''(\vec{q},\w)$ as in Fig.~\ref{fig:spec1},
but for a density of $p$ {\em strong} intradimer couplings with $J_2 = 3J_1/2$.
} \label{fig:bimjpe2}
\end{figure*}

We now turn to a detailed analysis of the weak substitution-induced magnetic order.
Fig.~\ref{fig:magwf} portraits a single realization of disorder, by showing the spatial
distribution of $J$ and the local magnetization $M_i$ as well as the wavefunctions for
selected eigenmodes of $\mathcal{H}_2$. Several features are apparent:
(i) The system develops islands of non-zero staggered magnetization,
Fig.~\ref{fig:magwf}(b), with a characteristic length scale of $\xi$, the correlation
length of the clean reference system. These islands, being rare events in the sense of
Griffiths \cite{griffiths}, exist in regions with larger concentration of weak interlayer bonds.
(ii) The distribution of local magnetization values becomes broad on logarithmic
scales, Fig.~\ref{fig:magwf}(f), a typical fingerprint of Griffiths behavior.
(iii) The low-energy ``magnon'' modes appear strongly localized on individual
magnetization islands, Fig.~\ref{fig:magwf}(c,d). This behavior persists for essentially
all energies below the gap of the clean reference system, whereas higher-energy modes
tend to be delocalized, Fig.~\ref{fig:magwf}(e).
We have confirmed this localization tendency by analyzing the inverse participation ratio
\cite{suppl}.

Taken together, this implies an interesting ``dual'' nature of the weakly ordered phase:
It displays a sharp Bragg peak in the static structure factor $S(\vec{q})$ at
$\vec{Q}=(\pi,\pi)$, Fig.~\ref{fig:stfac1}(a), as the underlying order is perfectly
staggered (albeit strongly inhomogeneous). At the same time, the small-$\w$ dynamic
response $\chi''(\vec{q},\w)$ is anomalously broad in $q$-space, due to the
disorder-induced mode localization. This is shown in Fig.~\ref{fig:stfac1}(b), where a
spin-wave-like two-peak structure is only visible for extremely small $\w/J\lesssim
0.02$.

Related localization and broadening phenomena have been observed and discussed both
theoretically and experimentally for a variety of random magnets, mainly inside
magnetically ordered \cite{cowley72,nagler84,uemura87,cherny02,mucciolo04,bouzerar10,cherny12}
or quantum-disordered phases \cite{aeppli00,bonca06}.
Our results show that, due to Griffiths physics, low-energy localization tendencies are
significantly stronger near quantum criticality than deep inside stable phases; for an
extended discussion see Ref.~\onlinecite{suppl}.

%%%%%%%%%%%%%%%%%%%%%%%%%%%%%%%%%%%%%%%%%%%%%%%%%%%%%%%%%%%%%%%%%%%%%%%

{\it Further results.}
We have studied other types of disorder, with selected results in
Fig.~\ref{fig:bimjpe2} and the supplemental material \cite{suppl}. Depending on the type
of disorder, the overall triplon bandwidth may either increase or decrease with doping;
the same applies to the apparent gap in  $\chi''(\w)$. For example, few {\em strong}
intradimer bonds decrease the width of main band, Fig.~\ref{fig:bimjpe2}, because
coherent triplon hopping is preempted through the strong bonds.
Band splitting like in Fig.~\ref{fig:spec1}(e) is visible for sizeable bimodal disorder;
for continuous distributions this is replaced by strong smearing of intensity.
Most importantly, spectral broadening at low energies appears generic near the QPT; this
effect is strongest if the dopants tend to drive the system towards the ordered phase, as
in Fig.~\ref{fig:spec1}.

We have also performed calculations for other 2d unfrustrated coupled-dimer models,
with qualitatively similar results which thus appear generic.

%%%%%%%%%%%%%%%%%%%%%%%%%%%%%%%%%%%%%%%%%%%%%%%%%%%%%%%%%%%%%%%%%%%%%%%

{\it Comparison to experiments.}
Recent experiments \cite{stone10,zheludev12,keimer12} have studied bond randomness
by ligand doping in spin-ladder materials. INS shows excitations with a larger energy
width and an increased spin gap as compared to the undoped case. This consistent with our
results for dopants which create locally stronger intradimer couplings, as in
Fig.~\ref{fig:bimjpe2}, or weaker interdimer couplings or both. For the investigated
materials, such properties can indeed be deduced from studies of the end members of the
doping series \cite{suppl}.

While INS data of bond-disordered nearly-critical dimer magnets in 2d or 3d are not
available to our knowledge, it is interesting to connect our results to \lsco: At
$x=0.145$ this compound displays a field-driven magnetic ordering transition, where INS
data indicate the closing of a spin gap, but in the presence of a non-diverging
correlation length \cite{chang09}. Since it is known that \lsco\ shows spatially disordered
charge stripes \cite{mv09} which in turn lead to modulated magnetic couplings, our
modelling of random-mass disorder can be expected to qualitatively describe the soft-mode
behavior of \lsco. Therefore we propose that quenched disorder tends to localize
low-energy magnetic modes at the QPT in \lsco, thus providing a cutoff to the apparent
magnetic correlation length. Disorder is also expected to cut-off one-parameter scaling
in $\chi''(\w,T)$ -- this is testable in future experiments.

%%%%%%%%%%%%%%%%%%%%%%%%%%%%%%%%%%%%%%%%%%%%%%%%%%%%%%%%%%%%%%%%%%%%%%%

{\it Summary.}
For coupled-dimer magnets, we have proposed an efficient method to calculate
real-frequency excitation spectra in the presence bond randomness across the entire phase
diagram. This allows one in particular to study magnetic excitations near quantum
criticality, where the effect of disorder is generically strong.
Using the bilayer Heisenberg model as an example, we have studied disorder-induced spectral weight
transfer and weak magnetic order. We find strong broadening of
low-energy excitation spectra due the localization tendency of the relevant modes.
Further high-resolution INS experiments, e.g., on Tl$_{1-x}$K$_x$CuCl$_3$ \cite{tlcudis},
which could test our predictions are called for.

Our method can be extended to the case with magnetic field, in order to access
excitations of disordered magnon Bose condensates, Bose glasses, or disordered
supersolids. A generalization to frustrated dimer lattices and to incommensurate order
is possible as well.

%%%%%%%%%%%%%%%%%%%%%%%%%%%%%%%%%%%%%%%%%%%%%%%%%%%%%%%%%%%%%%%%%%%%%%%

% \acknowledgments

I thank E. Andrade, S. Burdin, J. Chang, R. Doretto, D. Joshi, M. M\"uller, C. R\"uegg,
and S. Wessel for discussions.
This research was supported by the DFG through FOR 960 and GRK 1621. In addition,
financial support by the Heinrich-Hertz-Stiftung NRW and the hospitality of the Centro
Atomico Bariloche during early stages of this project are gratefully acknowledged.

%%%%%%%%%%%%%%%%%%%%%%%%%%%%%%%%%%%%%%%%%%%%%%%%%%%%%%%%%%%%%%%%%%%%%%%

\end{document}

\title{
Supplementary information for:\\
Excitation spectra of disordered dimer magnets near quantum criticality
}

\author{Matthias Vojta}
\affiliation{Institut f\"ur Theoretische Physik, Technische Universit\"at Dresden,
01062 Dresden, Germany}
\date{Aug 16, 2013}

\maketitle

%%%%%%%%%%%%%%%%%%%%%%%%%%%%%%%%%%%%%%%%%%%%%%%%%%%%%%%%%%%%%%%%%%%%%%%

\section{Bond operators with disorder: Numerics}

In the following we describe details of our bond-operator approach to dimer magnets
with quenched disorder, in particular the numerical implementation. We work with
finite-size systems of $N=L^2$ dimers on a square lattice with periodic boundary
conditions; the generalization to other lattices is straightforward.

\subsection{Optimum product state}

The basis rotation in the four-dimensional Hilbert space of each dimer, Eq.~\eqtrafo\ in
the main text, is determined by the requirement that $|\tilde{\psi}_0\rangle=\prod_i
|\tilde{t}_0\rangle_i$ is the best product-state approximation to the ground state of
$\mathcal{H}$.
The state $|\tilde{\psi}_0\rangle$ can be obtained in various ways: (i) by minimizing
$\mathcal{H}_0 = \langle\tilde{\psi}_0|\mathcal{H}|\tilde{\psi}_0\rangle$, (ii) by
demanding $\mathcal{H}_1$ to vanish, as the latter would create a condensate of
$\tilde{t}_\alpha$ bosons, and (iii) dimer mean-field theory. It is easy to
see that these methods are equivalent: First, upon minimizing $\mathcal{H}_0$ one finds
a saddle point where by construction linear variations (and thus $\mathcal{H}_1$) vanish.
Second, the minimum condition can be converted into mean-field theory as follows.
The variation of $\langle\tilde{\psi}_0|\mathcal{H}|\tilde{\psi}_0\rangle$ w.r.t.
$|\tilde{t}_0\rangle_i$ or, equivalently w.r.t. $U_{0k'}^{(i)}$ reads
\begin{equation}
\label{mincond}
\frac{\delta \langle\mathcal{H}\rangle} {\delta U_{0k'}^{(i)}}
=
J_i \frac{\delta \langle \vec{S}_{i1} \cdot \vec{S}_{i2} \rangle_i } {\delta U_{0k'}^{(i)}} +
\sum_m \frac{\delta \langle \vec{S}_{im} \rangle_i} {\delta
U_{0k'}^{(i)}}\cdot \vec{h}_{im} = 0
\end{equation}
where $\langle \ldots \rangle_i$ is a shorthand for ${\phantom .}_i\langle\tilde{t}_0|
\ldots|\tilde{t}_0\rangle_i$ and $\vec{h}_{im} = \sum_{i'm'} K_{ii'}^{mm'}
\langle\vec{S}_{i'm'}\rangle_{i'}$. The condition Eq.~\eqref{mincond} is the same as the
minimum condition for the single-dimer mean-field Hamiltonian
\begin{equation}
\label{hmf1}
\mathcal{H}_{\rm MF}^{(i)} = J_i \vec{S}_{i1} \cdot \vec{S}_{i2} +
\sum_{i'mm'} K_{ii'}^{mm'} \vec{S}_{im} \cdot \langle\vec{S}_{i'm'}\rangle_{i'}.
\end{equation}
For this single-dimer problem the variation w.r.t. $|\tilde{t}_0\rangle_i$ is a variation
in the full Hilbert space and hence yields the exact ground state. This establishes
the equivalence of the three criteria. We also note that $|\tilde{\psi}_0\rangle$ becomes
the exact ground state of $\mathcal{H}$ in the limit of infinite coordination number
between the dimers, see Sec.~\ref{sec:beyond} below.

To practically determine the optimum collinearly ordered product state for the
Hamiltonian $\mathcal{H}$ with a given realization of disorder, i.e. given random
couplings, we have found it most efficient to iterate the coupled set of $N$ mean-field
problems in Eq.~\eqref{hmf1} until convergence:
We start with an initial collinear guess for the $\langle\vec{S}_{im}\rangle$. Then,
diagonalizing the $4\times 4$ Hamiltonian $\mathcal{H}_{\rm MF}^{(i)}$ in Eq.~\eqref{hmf1} for
each dimer site $i$
gives a $4\times 4$ matrix of eigenvectors which we choose to be $U^{(i)}$. In
particular, the lowest-energy eigenvector $U_{0k'}^{(i)}$ can be used to calculate new
$T=0$ values of the $\langle\vec{S}_{im}\rangle$, which closes the self-consistency
loop.

\subsection{Conversion of the Hamiltonian}

To calculate the excitation spectrum, the Hamiltonian $\mathcal{H}$ needs to be expressed in terms of
the local $\tilde{t}_{i\alpha}$ excitation operators, as described in the following.

The starting point is the representation of the spins $\vec{S}_{im}$ via transition operators
between the states $|t_k\rangle_i$ of a dimer,
\begin{equation}
\label{strans}
{S}_{im}^\alpha = \sum_{kk'} s_{kk'}^{\alpha m} |t_k\rangle_i {\phantom .}_i\langle t_{k'}|
\end{equation}
with $4\times4$ matrices $s^{\alpha m}$ for the spin components $S^\alpha$ ($\alpha =
x,y,z\equiv 1,2,3$) of the $m=1,2$ spins:
\begin{eqnarray}
s^{x1,2} &=& \frac{1}{2} \left(
\begin{array}{cccc}
0 & \pm 1 & 0 & 0 \\
\pm 1 & 0 & 0 & 0 \\
0 & 0 & 0 & -i \\
0 & 0 & i & 0
\end{array}
\right), \nonumber\\
s^{y1,2} &=& \frac{1}{2} \left(
\begin{array}{cccc}
0 & 0 & \pm 1 & 0 \\
0 & 0 & 0 & i \\
\pm 1 & 0 & 0 & 0 \\
0 & -i & 0 & 0
\end{array}
\right), \nonumber\\
s^{z1,2} &=& \frac{1}{2} \left(
\begin{array}{cccc}
0 & 0 & 0 & \pm 1 \\
0 & 0 & -i & 0 \\
0 & i & 0 & 0 \\
\pm 1 & 0 & 0 & 0
\end{array}
\right).
\label{smat}
\end{eqnarray}
This is of course equivalent to the bond-operator representation of Sachdev and Bhatt,\cite{bondop}
\begin{eqnarray}
S_{i1,2}^\alpha &=& \frac{1}{2}\left(\pm t_{i0}^\dagger t^{\phantom{\dagger}}_{i\alpha}
              \pm t^\dagger_{i\alpha} t_{i0}
          - i\epsilon_{\alpha\beta\gamma}t^\dagger_{i\beta} t^{\phantom{\dagger}}_{i\gamma} \right),
\label{spin-bondop}
\end{eqnarray}
in the Hilbert space with $\sum_k  t^\dagger_{ik} t_{ik} =1$ bosons.
After the basis rotation \eqtrafo, Eq.~\eqref{strans} becomes
\begin{equation}
\label{strans2}
{S}_{im}^\alpha = \sum_{kk'} \tilde{s}_{i,kk'}^{\alpha m} |\tilde{t}_k\rangle_i {\phantom .}_i\langle \tilde{t}_{k'}|
\end{equation}
with the transformed spin matrices now being site-dependent:
\begin{equation}
\label{smat2}
\tilde{s}_{i,kk'}^{\alpha m} = \sum_{ll'} (U^\dagger)_{lk}^{(i)} s_{ll'}^{\alpha m} U_{k'l'}^{(i)}.
\end{equation}

The full Hamiltonian in terms of transition operators takes the (exact) form
\begin{eqnarray}
\mathcal{H}
&=& \sum_{ikk'} X_i^{kk'} |\tilde{t}_{k}\rangle_i {\phantom .}_i\langle \tilde{t}_{k'}| \nonumber\\
&+& \sum_{ii'kk'll'} Y_{ii'}^{kk'll'} |\tilde{t}_{k}\rangle_i |\tilde{t}_{l}\rangle_{i'}
 {\phantom
.}_i\langle \tilde{t}_{k'}| {\phantom
.}_{i'}\langle \tilde{t}_{l'}|.
\end{eqnarray}
Its coefficients $X_i^{kk'}$ and $Y_{ii'}^{kk'll'}$ are related to the original couplings
$J_i$ and $K_{ii'}^{mm'}$, respectively. They can be generated numerically for arbitrary
$U_{kk'}^{(i)}$ of a finite-size system with a given realization of disorder, using the
equations \eqh, \eqref{strans2}, and \eqref{smat2}.

This Hamiltonian is now rewritten into $\tilde{t}_{i\alpha}$ ($\alpha=1,2,3$) bosonic
operators which are defined to act on the background state $\prod_i
|\tilde{t}_0\rangle_i$, i.e., $\tilde{t}_{i\alpha}^\dagger|\tilde{t}_0\rangle_i =
|\tilde{t}_\alpha\rangle_i$.
Using the decomposition $\mathcal{H} = \mathcal{H}_0 + \mathcal{H}_1 + \mathcal{H}_2 + \mathcal{H}_3 +
\mathcal{H}_4$ we have
\begin {widetext}
\begin{eqnarray}
\mathcal{H}_0 &=& \sum_i X_i^{00} + \sum_{ii'} Y_{ii'}^{0000}, \nonumber\\
\mathcal{H}_1 &=& \sum_{i\alpha} \left(X_i^{0\alpha} \tilde{t}_{i\alpha} + X_i^{\alpha 0}
\tilde{t}_{i\alpha}^\dagger \right) + \sum_{ii'\alpha} \left(Y_{ii'}^{0\alpha00} \tilde{t}_{i\alpha} + Y_{ii'}^{000\alpha} \tilde{t}_{i'\alpha} + Y_{ii'}^{\alpha 000}
\tilde{t}_{i\alpha}^\dagger + Y_{ii'}^{00\alpha 0}
\tilde{t}_{i'\alpha}^\dagger \right), \nonumber\\
\mathcal{H}_2 &=& \sum_{i\alpha\beta} (X_i^{\alpha\beta}-X_i^{00}\delta_{\alpha\beta}) \tilde{t}_{i\alpha}^\dagger \tilde{t}_{i\beta}
+\sum_{ii'\alpha\beta} \left[ (Y_{ii'}^{\alpha\beta00}-Y_{ii'}^{0000}\delta_{\alpha\beta}) \tilde{t}_{i\alpha}^\dagger \tilde{t}_{i\beta}
 (Y_{ii'}^{00\alpha\beta}-Y_{ii'}^{0000}\delta_{\alpha\beta}) \tilde{t}_{i'\alpha}^\dagger \tilde{t}_{i'\beta}\nonumber \right] \\
&+&
\sum_{ii'\alpha\beta} \left( Y_{ii'}^{\alpha00\beta} \tilde{t}_{i\alpha}^\dagger \tilde{t}_{i'\beta} +
Y_{ii'}^{0\beta\alpha 0} \tilde{t}_{i'\alpha}^\dagger \tilde{t}_{i\beta} +
Y_{ii'}^{\alpha0\beta0} \tilde{t}_{i\alpha}^\dagger \tilde{t}_{i'\beta}^\dagger +
Y_{ii'}^{0\alpha0\beta} \tilde{t}_{i\alpha} \tilde{t}_{i'\beta}
\right)
\label{h012}
\end{eqnarray}
\end{widetext}
where $\alpha,\beta=1,2,3$.
Similar expressions can be found for $\mathcal{H}_3$ and $\mathcal{H}_4$ which are, however, not
needed in the following. In each run, we have checked that $\mathcal{H}_1$ vanishes to numerical
accuracy provided that the iteration of the product state has converged.

\subsection{Gaussian fluctuations and susceptibility}

The Gaussian fluctuations of interest are contained in $\mathcal{H}_2$, which
describes $3N$ modes where $N$ is the number of dimers, and the factor of three arises
from the spin polarizations $\alpha=x,y,z$.

We generate a matrix form of $\mathcal{H}_2$ numerically from the coefficients
$X_i^{kk'}$ and $Y_{ii'}^{kk'll'}$ via Eq.~\eqref{h012}. In both the paramagnetic and the
collinearly ordered cases studied here, the polarization directions decouple, such that
$\mathcal{H}_2$ is block-diagonal.
Then, it remains to perform a bosonic Bogoliubov transformation for the
transverse and longitudinal fluctuations separately. Each amounts to the numerical
diagonalization of a $2N \times 2N$ non-Hermitean matrix. We have implemented the
procedure described in the appendix of Ref.~\onlinecite{wessel} using standard LAPACK
routines. An efficient implementation can be achieved using the OpenMP version of Intel's
MKL library, which allowed us to access system sizes up to $N=128^2$ in the collinear
case using a desktop computer.

Each diagonalization yields (per polarization) $2N$ eigenvectors and eigenvalues, with
the latter coming in pairs with opposite sign. In the presence of a symmetry-breaking
condensate, we always find at least one pair of eigenvalues with magnitude smaller than
$10^{-5}J$ in the transverse piece of $\mathcal{H}_2$, corresponding to the expected
zero-energy (Goldstone) mode.

The positive-energy eigenvalues $\omega_n$ and their eigenvectors $(u_n,v_n)$ describe
the Gaussian fluctuation modes $\tau_n$ of $\mathcal{H}$, which are linear combinations of
the $\tilde{t}$ operators:
\begin{equation}
\label{taun}
\tau_n = \sum_{i\alpha} \left(u_{ni\alpha} \tilde{t}_{i\alpha} + v_{ni\alpha} \tilde{t}_{i\alpha}^\dagger\right).
\end{equation}
The $\omega_n$ and $(u_n,v_n)$ can be used to extract physical observables. In the
paper, we have concentrated on the dynamic spin susceptbility at $T=0$ in the
single-magnon approximation
\begin{equation}
\label{chieq}
\chi''_{\alpha}(\vec{q},\w) =
\left[\sum_{n}\left|M_{n}^{\alpha}(\vec{q})\right|^{2}\delta\left(\w-\w_{n}\right)\right]_{\rm
av}
\end{equation}
where $[\,]_{\rm av}$ denotes disorder averaging, and $M_{n}^{\alpha}(\vec{q})$ is the
matrix element $\langle0|S^{\alpha}(\vec{q})|n\rangle$ of the
Fourier-transformed spin operator (in the single-magnon approximation),
\begin{equation}
S^{\alpha}(\vec{q}) = \sum_{im} e^{i \vec{q}\cdot\vec{r}_{im}}
\sum_{\beta} \left(\tilde{s}_{i\beta 0}^{\alpha m} \tilde{t}_{i\beta}^\dagger + \tilde{s}_{i0 \beta}^{\alpha m}
\tilde{t}_{i\beta}\right)
\end{equation}
with the eigenmode $|n\rangle = \tau_n^\dagger|0\rangle$. Here, $\vec{r}_{im}$ is the
lattice coordinate of an individual spin.
For the bilayer model under consideration, $\vec{q}$ has a $z$
component $q_z=0,\pi$; we restrict our attention to the staggered response,
$q_z=\pi$, because the low-energy modes have significant weight only in this channel.

For a $L\times L$ system, $\chi''_{\alpha}(\vec{q},\w)$ can then be calculated with a
momentum resolution in $(q_x,q_y)$ of $L$ points per direction. The delta function in Eq.~\eqref{chieq}
may be converted into a Lorentzian with artificial width. However, for momentum-resolved
spectra we found it advantageous to use binning instead, because the exact intensity in
the clean case diverges as $1/\w$ upon approaching $\vec Q$ for transverse (Goldstone)
modes, such that a Lorentzian broadening leads to unphysical weight at elevated energies.

In order to improve the momentum resolution, we have also implemented twisted boundary
conditions, or equivalently, supercells: The system is treated as a periodic arrangement
of $M\times M$ subsystems of size $L\times L$ each and overall periodic boundary
conditions. Then, $M^2$ diagonalizations of a system with size $L^2$ are required, and
the number of momentum points per direction is $M\times L$. We note that supercell
calculations render the Hamiltonian matrix inevitably complex, while it is otherwise real
in the collinear case considered here. Most results in the paper have been obtained for
$L=64$ and $M=2$, and we show the transverse piece of $\chi''$ only.

We note that the present harmonic (or Gaussian) approximation for the problem with
disorder is designed to capture disorder-induced mode broadening, but does not describe
interaction-induced mode broadening, see also Sec.~\ref{sec:beyond} below. Phase-space
arguments imply that the former dominates at low energies whereas the latter can be
expected to become important at elevated energies.

\subsection{Static structure factor}

The static spin structure factor is defined as
\begin{equation}
S(\vec{q}) = \frac{1}{2N}
\left[\sum_{ii'mm'} \langle \vec{S}_{im} \cdot \vec{S}_{i'm'} \rangle e^{-i\vec{q}\cdot\vec{r}_{ii'mm'}}\right ]_{\rm av}
\end{equation}
with $2N$ the total number of spins and $\vec{r}_{ii'mm'} = \vec{r}_{im}-\vec{r}_{i'm'}$
the distance between a pair of spins.
In the bilayer model, the ordered state displays a Bragg peak at $\vec{Q}=(\pi,\pi,\pi)$;
for $N\to\infty$ its intensity is related to the staggered magnetization per spin via
$M_s^2 = S(\vec{Q})/(2N)$.

In Fig.~4 of the main text we show $S(\vec q)$ for $q_z=\pi$; for the finite-size
scaling $S(\vec{q})$ has been calculated in a mean-field approximation only, $\langle
\vec{S}_{im} \cdot \vec{S}_{i'm'} \rangle \rightarrow \langle \vec{S}_{im}\rangle \cdot
\langle\vec{S}_{i'm'} \rangle$. For small systems we have checked that the fluctuation
corrections only lead to minor quantitative changes.

%%%%%%%%%%%%%%%%%%%%%%%%%%%%%%%%%%%%%%%%%%%%%%%%%%%%%%%%%%%%%%%%%%%%%%%%%%%%%%%%%%%%%%%%%%%%%

\section{Bond operators for the clean bilayer magnet}

\begin{figure}[!t]
\includegraphics[width=0.4\textwidth]{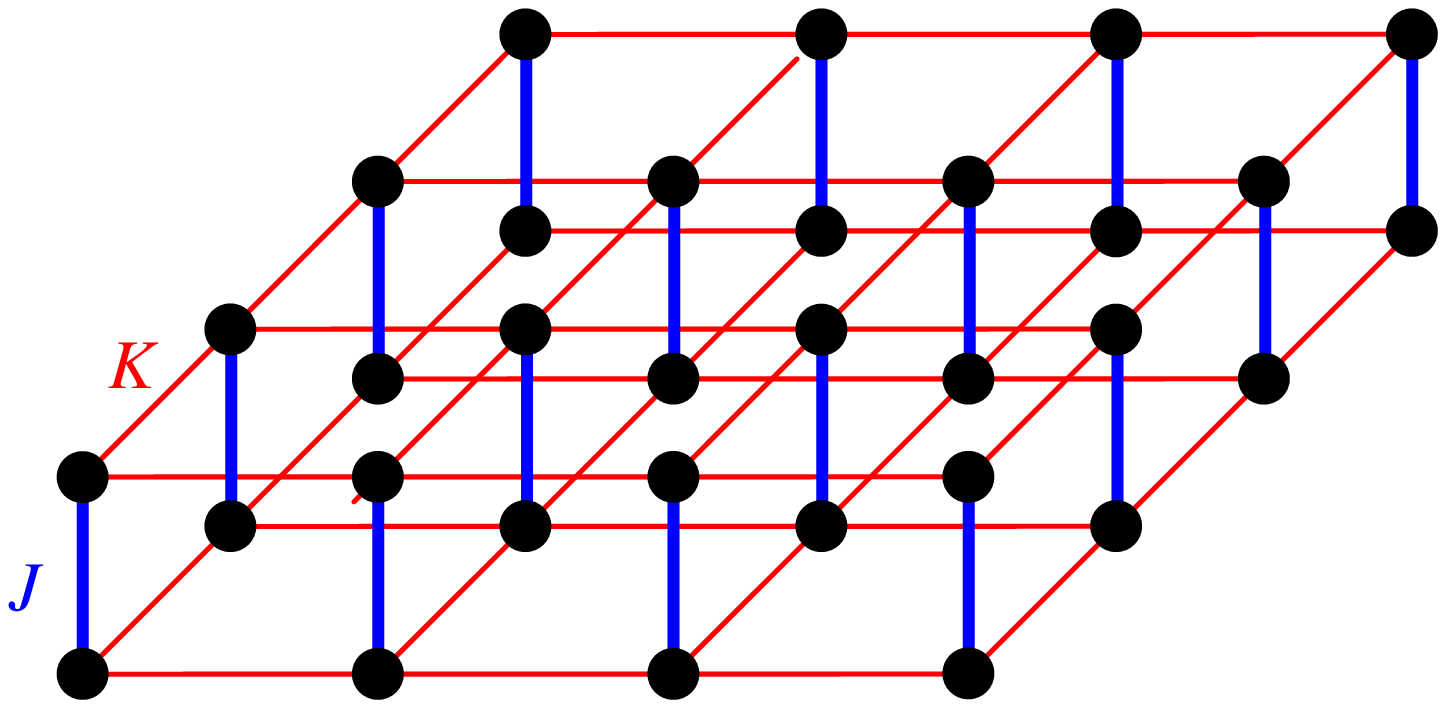}
\caption{%(color online)
Square-lattice bilayer Heisenberg model with interlayer coupling $J$ and intralayer coupling $K$.
} \label{fig:bilayer}
\end{figure}

To make the presentation self-contained, we quickly summarize the most relevant results
of the bond-operator approach to the bilayer Heisenberg magnet, Fig.~\ref{fig:bilayer},
without quenched disorder.

\subsection{Harmonic theory: Non-interacting bosons}

A number of papers have applied versions of the bond-operator formalism to the bilayer
Heisenberg model. In many cases, interactions between triplet excitations are neglected,
i.e., only bilinear terms in the bond-operator Hamiltonian are kept. The approaches
differ in the treatment of the Hilbert-space constraint: It is either fulfilled on
average, leading to so-called bond-operator mean-field
theory,\cite{bondop,gelfand98,yu99,ganesh11} or ignored completely (at the quadratic
level) in the spirit of linear spin-wave theory.\cite{kotov,sommer}
Here we focus on Ref.~\onlinecite{sommer} which follows the latter concept, as this
allows one to consistently treat the ordered phase across the entire phase diagram such
that Goldstone's theorem is automatically fulfilled. We also restrict ourselves to the
case of zero applied field.

The basis transformation in Eq.~(2) in the main text takes the following simple form:
\begin{equation}
\tilde{t}_{i0}^\dagger = \frac{t_{i0}^\dagger + \lambda e^{i\vec{Q}\cdot\vec{R}_i} t_{i3}^\dagger}{1+\lambda^2},~~
\tilde{t}_{i3}^\dagger = \frac{t_{i3}^\dagger - \lambda e^{i\vec{Q}\cdot\vec{R}_i} t_{i0}^\dagger}{1+\lambda^2}
\end{equation}
and $\tilde{t}_{i1}^\dagger =t_{i1}^\dagger$, $\tilde{t}_{i2}^\dagger =t_{i1}^\dagger$.
Here $\lambda$ is parameterizes the linear combination of singlet and $z$ triplet, and
$\vec{Q}=(\pi,\pi)$ is the ordering wavevector in the dimer lattice of the $\vec{R}_i$.
Minimization of the product-state energy, $\mathcal{H}_0 =
\langle\tilde{\psi}_0|\mathcal{H}|\tilde{\psi}_0\rangle$, yields
\begin{equation}
\lambda^2 = \frac{4K-J}{4K+J}.
\end{equation}

For $K/J < 1/4$ the system is paramagnetic, and from $\mathcal{H}_2$ one finds the
energy dispersion to be
\begin{eqnarray}
\label{cleandisp}
\w_{\vec q} &=& \sqrt{A_{\vec q}^2 - B_{\vec q}^2},\\
B_{\vec q} &=& 2K\gamma_{\vec q},~A_{\vec q} = J + B_{\vec q},\nonumber\\
\gamma_{\vec q} &=& (\cos q_x + \cos q_y)/2,
\end{eqnarray}
with a spin gap given by
\begin{equation}
\label{cleangap}
\Delta = \sqrt{J (J - 4K)}
\end{equation}
and a triplon bandwidth of
\begin{equation}
\label{cleanbw}
W = \sqrt{J} \left( \sqrt{J + 4K} - \sqrt{J - 4K} \right).
\end{equation}

For $K/J > 1/4$ the system is in an antiferromagnetic state, with a product-state
staggered magnetization
\begin{equation}
M_s = \frac{\lambda}{1+\lambda^2}
\end{equation}
per spin.
This value is somewhat reduced by Gaussian fluctuations; in particular in the limit of
decoupled layers, $J\to 0$, where $\lambda=1$ one recovers the spin-wave value $M_s =
0.66$ (Ref.~\onlinecite{sommer}).

\subsection{Beyond non-interacting bosons}
\label{sec:beyond}

\subsubsection{Microscopic treatment}

Going beyond the non-interacting (i.e. linearized) theory requires to take into account
the quartic interactions in $\mathcal{H}_4$ (cubic terms are absent from the bilayer
model) as well as the hard-core constraint.
An efficient approximation for the latter is given by the Brueckner theory, which yields
an accurate value for the location of the critical point, as shown by Kotov {\em et
al.}\cite{kotov} However, this approach cannot be easily generalized to the ordered
phase.
An earlier approach by Chubukov and Morr,\cite{Chub95} inspired by non-linear spin-wave
theory, works in both phases, but suffers from divergencies at higher orders, probably
because it lacks a small control parameter.

Notably, it can be shown that a systematic expansion can be constructed in the limit of
large coordination number $z$ (or, alternatively, large spatial dimension $d$), i.e.,
using $1/z$ (or $1/d$) as a small parameter.\cite{joshi} First, fluctuation corrections to the
product state $|\tilde{\psi}_0\rangle$ vanish as $z\to\infty$; this is similar in spirit
to spin-wave theory where fluctuation corrections vanish in the limit of large spin.
Second, the harmonic fluctuations described by non-interacting triplons constitute the
leading-order correction to the product state, i.e., their contribution to thermodynamic
quantities is of order $1/z$. Finally, higher-order corrections can be perturbatively
calculated order by order in $1/z$. The complexity is comparable to that of non-linear
spin-wave theory.\cite{joshi} For the present problem, with quenched disorder,
corrections beyond the harmonic approximation can be calculated in principle, but are
beyond the scope of this paper.

\subsubsection{Effective modelling}

In general, anharmonic effects contribute to thermodynamic quantities and modify spectral
properties. If one restricts the attention to the dispersion and spectral weight of the
triplon quasiparticles, then the effect of interactions may be captured in terms of
renormalized model parameters, i.e., the full mode dispersion of the original model can
be approximately represented as the dispersion of harmonic triplons in a renormalized
model. This has been investigated in some detail for spin ladders in
Ref.~\onlinecite{eder98}, where it was found that important spectral properties are well
reproduced in such a simplied approach when compared against a numerical exact
diagonalization of the original model.

For the QPT in the present bilayer model, this implies that the interaction-induced shift
of the critical point -- from $(J/K)_c=4$ to $(J/K)_c = 2.5220(1)$ (Ref.~\onlinecite{sandvik06}) --
can be captured, to leading order, in terms of renormalized $J$, $K$. In this paper, we
account for this fact in the simplest fashion, namely by specifying the parameter ratio
$J/K$ relative to the QCP location.

%%%%%%%%%%%%%%%%%%%%%%%%%%%%%%%%%%%%%%%%%%%%%%%%%%%%%%%%%%%%%%%%%%%%%%%%%%%%%%%%%%%%%%%%%%%%%

\begin{figure*}[!t]
\includegraphics[width=0.9\textwidth]{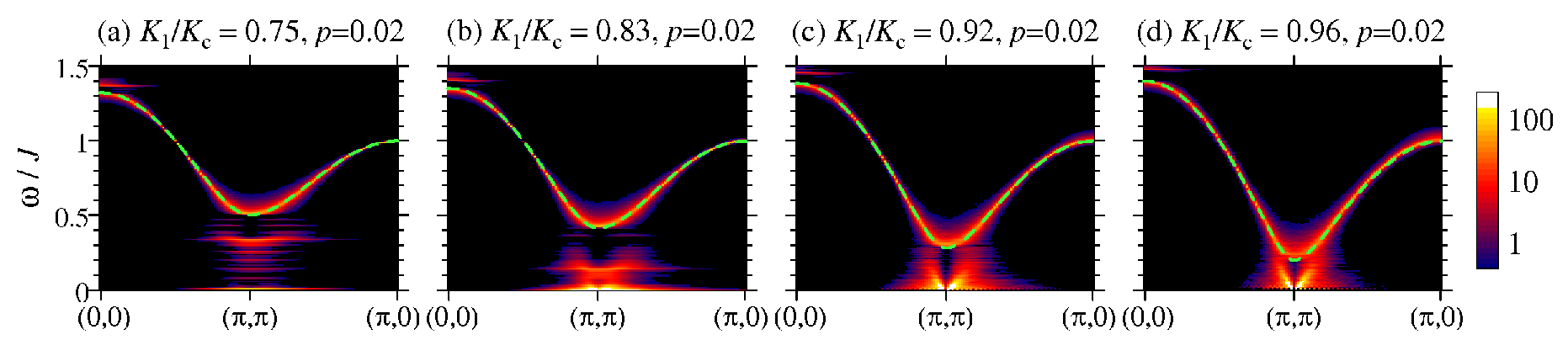}
\caption{%(color online)
Susceptibility $\chi''(\vec{q},\w)$, but now for disorder in the intralayer couplings $K$:
a density of $p$ dimers has strong intralayer couplings $K_2 = 2 K_1$ with all neighbors.
} \label{fig:bimjpca1}
\end{figure*}

\begin{figure*}[!t]
\includegraphics[width=0.9\textwidth]{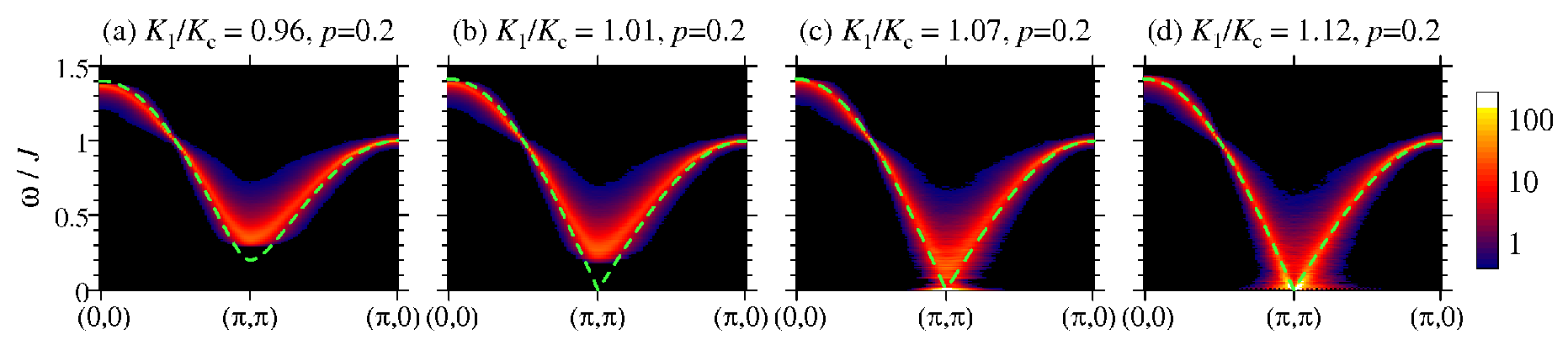}
\caption{%(color online)
Susceptibility $\chi''(\vec{q},\w)$ for disordered intralayer couplings $K$ as in Fig.~\ref{fig:bimjpca1},
but now a density of $p$ dimers has weak intralayer couplings with $K_2 = K_1 / 2$ with all neighbors.
} \label{fig:bimjpca2}
\end{figure*}

\section{Results for different types of disorder}

In the main paper, results were discussed for disordered interlayer (or intra-dimer)
couplings $J$, specifically a small concentration of weakened (Figs.~1-4) or strengthened
(Fig.~5) interlayer couplings. Here we first discuss some general aspects of disorder,
also with an eye towards real materials, and then present data for other types of
disorder.

\subsection{Griffiths regime and paramagnetic gap}

For the bounded disorder distributions considered in this paper, there is always a
well-defined (hard) excitation gap for sufficiently small $K/J$. Upon increasing $K$, the
gap closes at the boundary to the quantum Griffiths regime.\cite{tv06} In our
mean-field-based treatment, (weak) magnetic order sets in at this point; non-linear
fluctuation corrections would be required to obtain the correct physical behavior,
namely a magnetic phase transition inside the Griffiths regime. In other words, at the
level of our approximation a paramagnetic quantum Griffiths regime is absent, and the
entire Griffiths regime is magnetically ordered. Despite this, we can expect that the
properties of the excitations are captured correctly except at ultra-low energies.

Depending on the type of disorder, the ``apparent'' energy gap in $\chi''$ may be much larger
than the actual gap, namely if the intensity in $\chi''$ is very small for small $\w$
-- this happens if the low-energy excitations are carried by rare configurations (i.e.
rare clusters of small $J$ or large $K$). This effect can be seen in Fig.~2a of the main
paper, where a true gap is absent at $p=0.05$, but the susceptibility has an apparent gap of
$\approx 0.1J$.

\subsection{Disorder in experiments}
\label{sec:exp}

Apart from fundamental interest, specific models of disorder may be motivated by the
experimental situation. Here, bond disorder is introduced via ligand substitution, as e.g.
in Tl$_{1-x}$K$_x$CuCl$_3$,\cite{tlcudis,tlcudis2}
(C$_4$H$_{12}$N$_2$)Cu$_2$(Cl$_{1-x}$Br$_x$)$_6$,\cite{zheludev12,zheludev12b}
IPA-Cu(Cl$_x$Br$_{1-x}$)$_3$,\cite{ipadis,keimer12}
or (Hpip)$_2$CuCl$_{4(1-x)}$Br$_{4x}$,\cite{hpipdis}
such that bimodal disorder models appear appropriate.

In general, such substitutions influence both intra-dimer ($J$) and inter-dimer ($K$)
couplings; this can be deduced from the end members of the substitution series. For
instance, TlCuCl$_3$ and KCuCl$_3$ have drastically different spins gaps of 0.7 and
2.5\,meV, respectively, and fits to INS data results in values of $J=5.42$\,meV and
$K_{1,2,3}=(-0.47,-1.43,0.62)$\,meV for TlCuCl$_3$, in contrast to $J=4.29$\,meV and
$K_{1,2,3}=(-0.21,-0.34,0.37)$\,meV for KCuCl$_3$; here $K_{1,2,3}$ are the strongest
couplings in the three-dimensional dimer network of these materials.\cite{cavadini01}
Therefore, a substitution Tl$\rightarrow$K in TlCuCl$_3$ tends to increase the spin gap;
the same applies to Cl$\rightarrow$Br in (C$_4$H$_{12}$N$_2$)Cu$_2$Cl$_6$ and
(Hpip)$_2$CuCl$_4$.

It can be expected that partial substitution of ligands locally changes $J$ and/or $K$
couplings, but this happens in a correlated fashion: A single substituted ligand atom
influences numerous local bond lengths and angles, such that various couplings in the
vicinity of the substituent are modified -- this applies in particular to interdimer
couplings which typically result from multiple exchange paths. Therefore, models of
uncorrelated bimodal disorder are a strong simplification and cannot fully capture the
disorder physics of real materials. We have therefore, in addition to bimodal
uncorrelated disorder in $J$, also considered correlated disorder in $K$, see below.

\begin{figure*}[!t]
\includegraphics[width=0.9\textwidth]{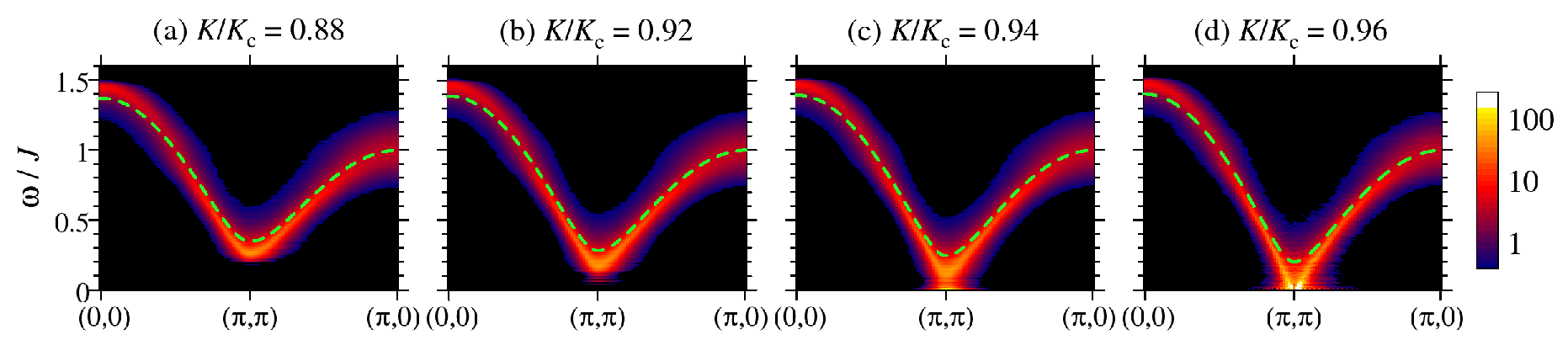}
\caption{%(color online)
Susceptibility $\chi''(\vec{q},\w)$, but now for a box distribution of interlayer
couplings $J$ with a width $\delta J = J/2$, i.e., the $J_i$ in Eq.~\eqh~ vary in the
interval $[0.75J,1.25J]$.
} \label{fig:boxjpe}
\end{figure*}

\begin{figure*}[!t]
\includegraphics[width=0.9\textwidth]{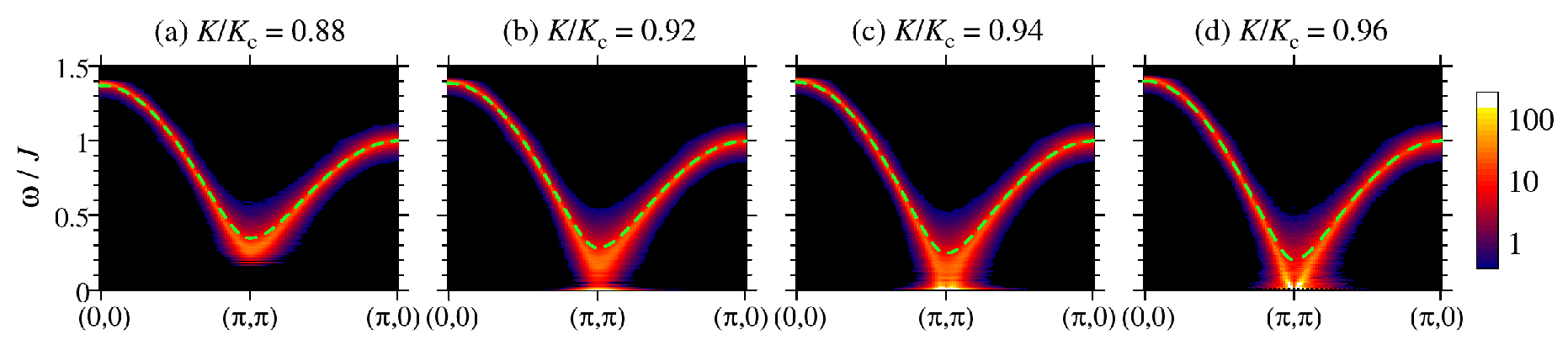}
\caption{%(color online)
Susceptibility $\chi''(\vec{q},\w)$, but now for a box distribution of intralayer
couplings $K$ with a width $\delta K = K$, i.e., the $K_{ii'}^{mm}$ in Eq.~\eqh\ vary in
the interval $[0.5K,1.5K]$.
} \label{fig:boxjpa}
\end{figure*}

\subsection{Bimodal bond disorder}

The case of disordered interlayer couplings $J$ was extensively discussed in the main
text. In Figs. 1-4 the focus was on a small concentration of weakened interlayer couplings.
Those have the tendency to reduce the spin gap by inducing in-gap states, which are
strongly localized and lead to broad modes at lowest energies near the transition
(Fig.~1).

The opposite case of a small concentration $p$ of stronger interlayer ($J$) couplings was
shown in Fig.~5 of the main paper. As a consequence of a few strong $J$ bonds, the
finite-$p$ excitation spectrum now shows a larger apparent gap as compared to the $p=0$
case, and ordering tendencies are suppressed. The few strong bonds induce non-dispersing
excitations above the main band and with weak intensity; at the same time the overall
width of the main band is reduced.
Near the transition, Fig.~5(c) there is still significant mode broadening
at low energies.

We now turn to disorder in the intralayer (or interdimer) couplings $K$. For uncorrelated
disorder in $K$, we have found that disorder effects are much weaker than that of
uncorrelated disorder in $J$, i.e., uncorrelated $K$ disorder efficiently averages out
because of the presence of many (here eight) $K$ bonds for each dimer.
Considering that uncorrelated disorder is unrealistic for real materials
(Sec.~\ref{sec:exp}), we have then chosen to model correlated disorder in the following
fashion: We randomly choose a (small) fraction $p$ of dimers, mimicking the location of
dopants, and then give {\em all} $K$ bonds emerging from such dimers the strength $K_2$,
while the remaining $K$ bonds have strength $K_1$. Results are shown in
Figs.~\ref{fig:bimjpca1} and \ref{fig:bimjpca2} for the cases of $p$ dimers with stronger
and weaker intralayer couplings, respectively.

Clearly, a small concentration of sites with stronger intralayer couplings,
Fig.~\ref{fig:bimjpca1}, has an effect similar to a small concentration of weaker
interlayer couplings, Fig.~1: The dopants introducde significant in-gap spectral weight
and drives the system towards the ordered state. The in-gap states themselves now show
significant energetic structure, arising from cluster of two, three, and more dimers
sites with stronger intradimer coupling.

In contrast, a small concentration of sites with weaker intralayer couplings,
Fig.~\ref{fig:bimjpca2}, leads to spectra similar to the case of a small concentration of
stronger interlayer couplings, Fig.~5 of the main paper, with bandwidth reduction and mode
broadening being the dominant effects.

\subsection{Box bond disorder}

For completeness, we have also considered uncorrelated continuous distributions of
coupling constants. Specifically, we have studied box distributions for either $J$ (with
non-disordered $K$) or $K$ (with non-disordered $J$). Corresponding results are shown in
Figs.~\ref{fig:boxjpe} and \ref{fig:boxjpa}, respectively.

Not surprisingly, box distributions tend to enlarge the effective bandwidth of magnetic
excitations, simply because the range of available couplings increases: For instance, a
variation of $J$ tends to shift the center energy of the triplon band, such that a system
with box-distributed $J$, Fig.~\ref{fig:boxjpe}, effectively samples bands corresponding
to a range of $J$ values.
Similarly, a variation of $K$ tends to modify the bandwidth, such that the bandwidth for
box-distributed $K\in[K-\delta K/2,K+\delta K/2]$ is larger than that corresponding to
the pure-$K$ case. However, the increase in bandwidth is moderate even for large $\delta
K$, Fig.~\ref{fig:boxjpa}, which demonstrates that uncorrelated disorder in the
intradimer couplings averages out as noted above.

Smearing of the excitations (but no splitting) is visible at all energies for both box
distributions, with smearing again being most pronounced at low energies near the
transition. In particular, strongly broadened low-energy modes are visible in
Figs.~\ref{fig:boxjpe}c and \ref{fig:boxjpa}c.

%%%%%%%%%%%%%%%%%%%%%%%%%%%%%%%%%%%%%%%%%%%%%%%%%%%%%%%%%%%%%%%%%%%%%%%

\section{Evolution of mode broadening and band width}

The calculated $\chi''(\vec{q},\w)$ enables a quantitative analysis of the evolution of
spectral properties with the disorder level, in particular the energetic mode broadening
and the triplon bandwidth.

\begin{figure}[!b]
\includegraphics[width=0.49\textwidth]{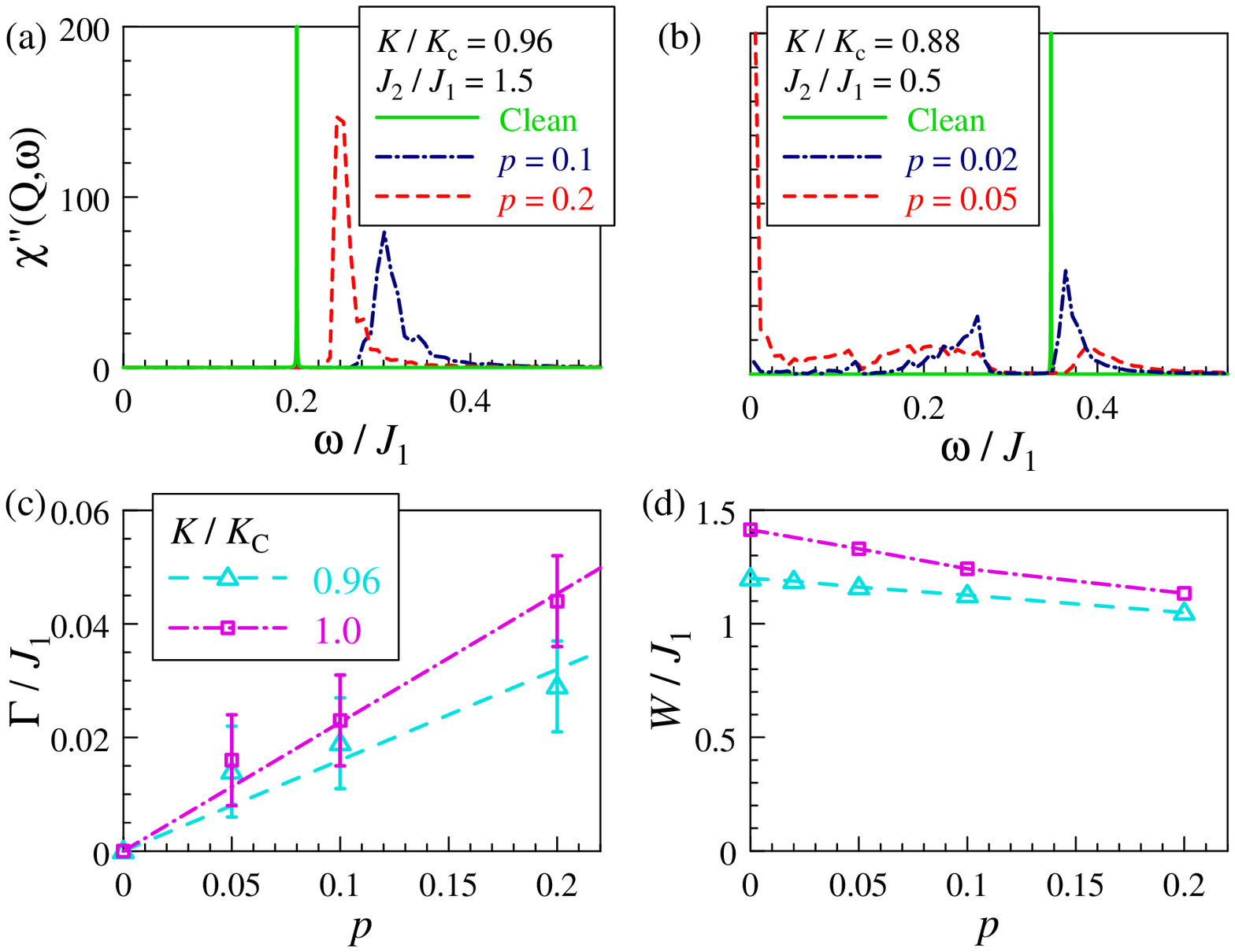}
\caption{%(color online)
(a) Excitation spectrum $\chi''(\vec{q},\w)$ at fixed ${\vec q} = {\vec Q} = (\pi,\pi)$,
for a small concentration of strong interlayer bonds with $J_2 = 3J_1/2$, as in
Fig.~5 of the main paper.
(b) Same as (a), but for a small concentration of weak interlayer bonds with $J_2 =
J_1/2$, as in Fig.~1 of the main paper.
(c,d) Linewidth at $\vec{Q}=(\pi,\pi)$ and triplon bandwidth for the case of of strong interlayer bonds with
$J_2 = 3J_1/2$ of function of $p$. The error bars arise primarily from the limited energy
resolution of our finite-size calculation; the lines are guide to the eye.
} \label{fig:lshape}
\end{figure}

A first impression can be obtained by considering the energy-dependent response at fixed
wavevector, as shown in Fig.~\ref{fig:lshape}(a,b) for the putative ordering wavevector
$\vec q = \vec Q = (\pi,\pi)$.
Both panels are for different disorder in the interlayer couplings: (a) has a small
concentration of {\em strong} interlayer bonds, which have the simple tendency to broaden
the mode and shift it to higher energies. In contrast, (b) has a small concentration of
{\em weak} interlayer bonds, which have two effects: The ``main'' mode is slightly
shifted upwards, but a significant amount of weight is created inside the gap.

In the case (a) of few strong interlayer bonds, it makes sense to quantify the mode
broadening. This is done in Fig.~\ref{fig:lshape}(c) for $\vec{q}=(\pi,\pi)$ and two
different values of $K/K_c$ corresponding to the data depicted in
Fig.~5(a,b) of the main paper. The data are consistent with a linewidth scaling linearly
with $p$, with a prefactor which increases upon approaching the QPT. (As noted before,
the broadening is largest in the regime where the apparent spin gap closes, see
Fig.~5(c).)

Fig.~\ref{fig:lshape}(d) displays the evolution of the magnon bandwidth for the same case
of few strong interlayer bonds; the small-intensity non-dispersing states at the upper
end of the spectrum, Fig.~5, have been ignored in this analysis. The bandwidth decreases
as expected from Eq.~\eqref{cleanbw}; the decrease is approximately linear for small $p$.
This qualitatively matches recent experimental INS data.\cite{keimer12,zheludev12b}

We finally note that non-linearities, i.e., triplon--triplon interactions, will also lead
to broadening and renormalization effects, which are not part of the present calculation.
However, interaction-induced broadening is typically small at low energies, i.e., near
the bottom of the excitation band, because of phase-space restrictions for particle
decay. Therefore, the disorder effects captured by our calculations can be expected to
dominate at low energies.

%%%%%%%%%%%%%%%%%%%%%%%%%%%%%%%%%%%%%%%%%%%%%%%%%%%%%%%%%%%%%%%%%%%%%%%

\section{Localization of low-energy modes}

As illustrated in Fig.~3(c,d) of the main text, the low-energy excitation modes of the
disordered coupled-dimer magnet tend to be spatially localized.
This can be quantified by considering the inverse participation ratio (IPR) of the
eigenmodes, with the general definition of ${\rm IPR} = \sum_i |\psi_i|^4$ for a
wavefunction $|\psi\rangle$, and $\psi_i = \langle i|\psi\rangle$ where $i$ is a lattice
site. The finite-size behavior of the IPR is indicative of localization: For spatially
extended states, the IPR scales with the system size as $1/N$, whereas it approaches a
constant for localized states. For exponential localization, the IPR value is then a
measure of the localization length, ${\rm IPR} \propto 1/\xi^d$.

For the eigenvectors $\tau_n$ of the bosonic Bogoliubov transformation \eqref{taun} we
define the IPR as
\begin{equation}
\label{ipr}
{\rm IPR}_n = \sum_{i\alpha} \left(|u_{n i\alpha}|^2 - |v_{n i\alpha}|^2\right)^2,
\end{equation}
note that $\sum_{i\alpha} (|u_{n i\alpha}|^2 - |v_{n i\alpha}|^2) = 1$.
Some previous papers\cite{wessel,mucciolo04} have employed an alternative definition,
\begin{equation}
\label{jpr}
{\rm IPR}'_n = \frac{\sum_{i\alpha} |v_{n i\alpha}|^4}{\sum_{i\alpha} |v_{n i\alpha}|^2};
\end{equation}
we have checked that both quantities show qualitatively similar properties.

\begin{figure}[!b]
\includegraphics[width=0.48\textwidth]{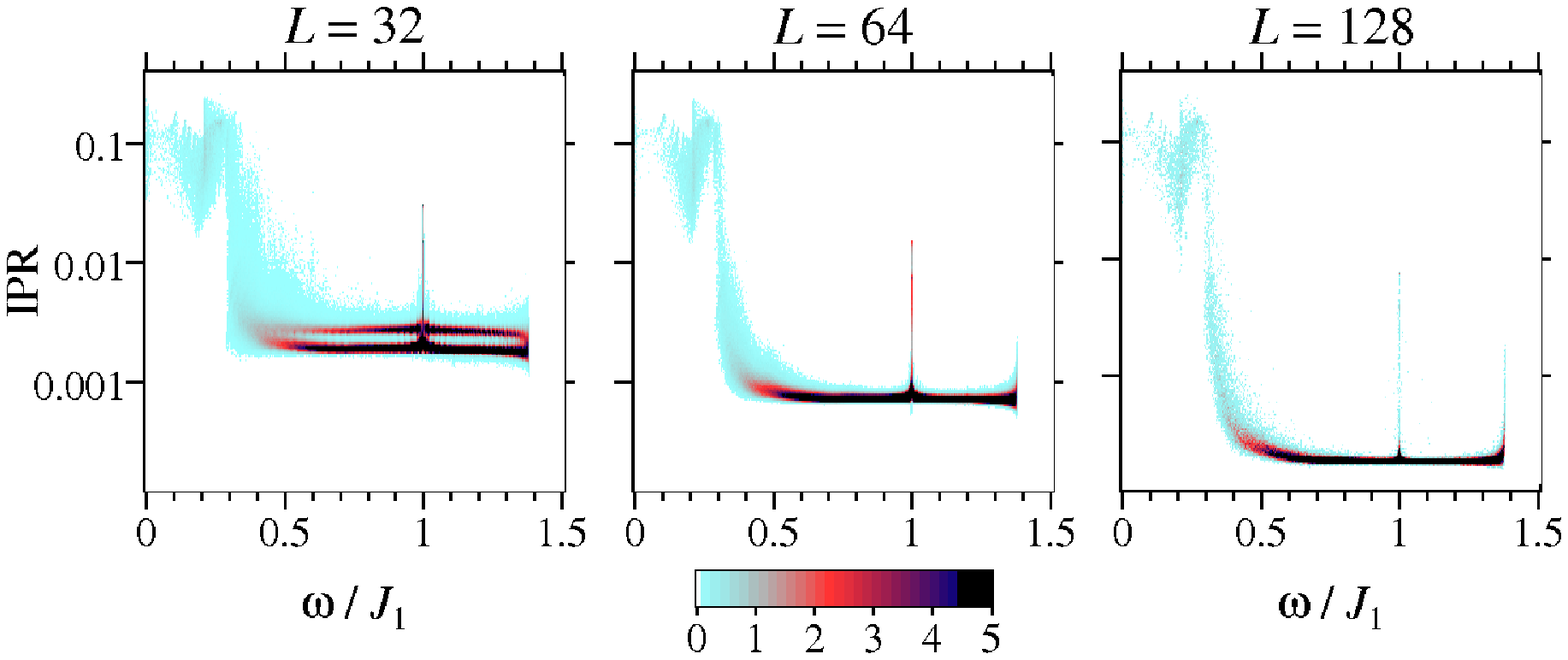}
\caption{%(color online)
Inverse participation ratio \eqref{ipr} for the eigenmodes of $\mathcal{H}_2$ of the
disordered bilayer Heisenberg model, with $K/K_c=0.92$ and $p=0.02$ weak interlayer
bonds with $J_2=J_1/2$ as in Fig.~1(g) of the main paper. The panels show the IPR as
function of the mode energy for different system sizes $L\times L$; the color encodes the
average number of modes (per IPR and energy interval and per disorder realization) in a
histogram fashion. Localization tendencies, with the IPR being independent of system
size, are pronounced for energies $\w<0.3J$, i.e., below the gap of the clean system.
} \label{fig:ipr}
\end{figure}

\begin{figure}[!t]
\includegraphics[width=0.48\textwidth]{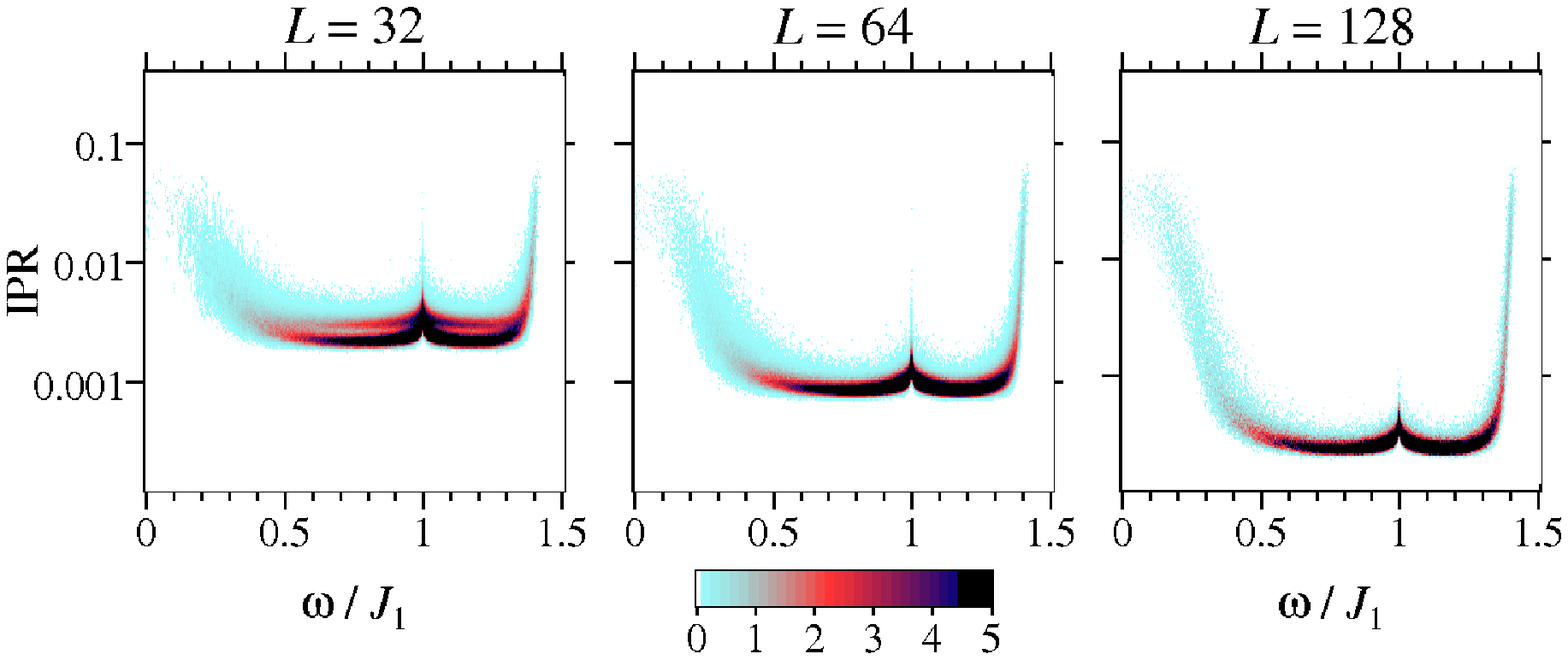}
\caption{%(color online)
Same as Fig.~\ref{fig:ipr}, but for a box distribution of intralayer couplings $K$ with
$K/K_c=0.92$ as in Fig.~\ref{fig:boxjpa}(b).
} \label{fig:ipr2}
\end{figure}

Energy-resolved histograms of IPR values \eqref{ipr} are shown in Fig.~\ref{fig:ipr} for
the same parameters as employed in Figs.~1(g), 3, 4 of the main text, i.e., $p=0.02$ weak
interlayer bonds. A scaling with inverse system size is visible for essentially all
energies $\w>0.35J$, indicating extended states, whereas states with $\w<0.3J$ are
clearly localized. Considering that the spin gap of the clean system is $\Delta=0.28J$,
it is obvious that it is the disorder-induced states below the gap which are subject to
localization. (The somewhat enhanced IPR at both $\w/J \approx 1$ and the upper band edge
arises primarily from the flat pieces of the mode dispersion.)
Comparing Figs.~1(g) and \ref{fig:ipr} also shows that energies with a large IPR correlate
with those showing a large broadening in momentum space, both indicative of a short
real-space localization length.

Fig.~\ref{fig:ipr2} illustrates similar localization tendencies for a box distribution of
couplings, albeit with a somewhat larger localization length. Here, localization takes
place a both band edges, but concerns a broader energy range near the band bottom as
compared to the top.

We conclude that disorder-induced localization of low-energy modes is generic near the QPT of
the bilayer Heisenberg model.

%%%%%%%%%%%%%%%%%%%%%%%%%%%%%%%%%%%%%%%%%%%%%%%%%%%%%%%%%%%%%%%%%%%%%%%

\section{Relation to broadening and localization phenomena in previous work on disordered magnets}

The physics of defects in magnetic insulators is an intensely studied subject. It has
long been recognized that defects modify the spectrum of magnetic excitations which can
be probed in optical, NMR, or neutron-scattering experiments.\cite{cowley72} Early works
mainly dealt with ordered magnets where, via chemical substitution, impurities with a
spin different from that of the host material were
introduced.\cite{cowley72,nagler84,uemura87} This includes the particularly simple case
of spinless impurities, a situation which has seen intense theoretical modelling over the
last decade.\cite{cherny02,mucciolo04,bouzerar10} In addition, spinless impurities
in quantum-disordered paramagnets were also studied.\cite{aeppli00,bonca06}

The present work differs in two main aspects from previous investigations of spectral
properties of disordered magnets:
First, we concentrate on systems in the vicinity of a magnetic quantum critical point, as
opposed to systems inside stable phases (either ordered or disordered). Quantum critical
systems are inherently unstable and thus more susceptible to disorder.
Second, we study a rather weak form of disorder, namely random bonds or,
technically speaking, random-mass disorder. Most earlier works instead dealt with
missing or substituted spins which induce random Berry phases. The latter form of
disorder is much stronger than random-mass disorder: for instance,
any small amount of spinless impurities brought into a confined gapped quantum paramagnet induces
zero-temperature magnetic order,\cite{imada97,yasuda01,wessel01,hase93,oosawa02}
whereas weak random-mass disorder leaves a gapped paramagnetic phase intact.

In terms of results for the excitation spectrum, $\chi''(\vec{q},\w)$,
these differences are visible in various respects:
(i) In our calculations, disorder effects are strongest near quantum criticality -- a
regime not covered by earlier work. In contrast, in ordered phases with Goldstone modes,
disorder tends to decouple from the excitations in the low-energy limit;\cite{gurarie03}
this sharpening of Goldstone modes is also seen in our results, see e.g. Figs. 1(e),
\ref{fig:bimjpca1}(d), \ref{fig:bimjpca2}(d), \ref{fig:boxjpe}(d), and
\ref{fig:boxjpa}(d).
(ii) Near criticality, we observe a large broadening already at small impurity
concentrations of 2\% (despite the fact that we consider random-mass disorder only). In
most previous work comparably large effects were present only for defect concentrations
of 10\% and above; notable exceptions are spinless impurities in certain spin-gap
magnets\cite{bonca06} and also in cuprates\cite{ybco} (with the latter being close to a magnetic
QPT).

Paranthetically, I note that a detailed analysis of lineshapes has not been performed
here (and is also missing from most previous works) -- this is left for future studies.

Finally, the aspect of spatial localization of the excitation modes deserves a
discussion. There is consensus that sufficiently strong disorder will induce mode
localization, and that weak disorder will not localize modes in high
dimensions.\cite{weichman08} However, localization properties of magnetic excitations in
low dimensions are not fully understood: While the standard arguments for Anderson
localization predict that all single-particle states are localized in generic
non-interacting 2d disordered systems, the following ingredients make the problem non-trivial:
(i) Magnons as well as triplons are hard-core bosons, i.e., their interactions are not
negligible. Understanding of localization properties of bosons in the presence of hard-core
interactions is an unsolved problem.
(ii) Even if interactions are neglected, the underlying (magnon or triplon)
single-particle Hamiltonian has a bosonic Bogoliubov-de-Gennes structure, such that the
standard arguments for localization do not directly apply. In addition, one has to
distinguish systems with and without Goldstone modes.\cite{gurarie03} Indeed, it has been
argued for certain 2d random magnets that delocalized spin-wave modes can exist, which
are energetically separated from localized modes by a mobility edge.\cite{laflo13}
Taking into account (i), this has to be considered speculative at present.

In diluted magnets near the percolation threshold, excitations at elevated energies have
been discussed in terms of fractons, i.e., localized modes which live on a percolating
network of sites.\cite{orbach94,ikeda94} This physics is not relevant for the cases of (weak)
bond disorder studied here, but it will be extremely interesting to consider excitations
subject to the interplay of both percolation and quantum criticality, as e.g. in the diluted
bilayer model studied in Ref.~\onlinecite{sandvik06b}.

%%%%%%%%%%%%%%%%%%%%%%%%%%%%%%%%%%%%%%%%%%%%%%%%%%%%%%%%%%%%%%%%%%%%%%%